# Hoki: Making BPASS accessible through Python

Heloise F. Stevance[1], J. J. Eldridge[1], and Elizabeth Stanway[2]

**1** Department of Physics, University of Auckland, New Zealand **2** Department of Physics and Astronomy, University of Warwick, United Kingdom

## Summary

We now know that a large number of stars are born in multiple systems. Additionally, more than 70% of massive stars are found in close binary systems, meaning that they will interact over the course of their lifetime (Sana et al., 2012). This has strong implications for their evolution as well as the transients (e.g supernovae) and the potential gravitational wave progenitors they produce. Therefore, in order to understand and correctly interpret astronomical observations of stellar populations, we must use theoretical models able to account for the effects of binary stars. This is the case of the Binary Population and Spectral Synthesis code (BPASS) (Eldridge et al., 2017; Stanway & Eldridge, 2018), which has been a staple of the field for over 10 years (Eldridge, Izzard, & Tout, 2008; Eldridge & Stanway, 2009). As is the case for most other theoretical models, the data products of BPASS are large, varied and complex. As a result, their use requires a level of expertise that is not immediately accessible to a wider community that may hold key observational data. The goal of `hoki` is to bridge the gap between observation and theory, by providing a set of tools to make BPASS data easily accessible and facilitate analysis. The use of Python is deliberate as it is a ubiquitous language within Astronomy. This allows BPASS results to be used naturally within the pre-existing workflow of most astronomers.

## Motivation

BPASS is a very powerful theoretical code which, given an Initial Mass Function (IMF), metallicity ($Z$) and age, can predict the Spectral Energy Distirbution (SED), transient rates, stellar numbers, yields and ionizing photon rates of a stellar population containing realistic fractions of binary stars (Eldridge et al., 2017; Stanway & Eldridge, 2018). These data products are made available to the community through the BPASS website and they are crucial to interpreting some astronomical observations. For example they have been used to better understand the rate of supernovae and their progenitors (Eldridge, Fraser, Smartt, Maund, & Crockett, 2013) as well as predicting the rate of black-hole mergers (Eldridge & Stanway, 2016). Although the BPASS outputs are available online, they can be challenging to use, due to the level of expertise required to perform the pre-processing necessary to allow optimal comparison to observational data. For example, creating synthetic Colour-Magnitude Diagrams (CMDs) requires searching through >100,000 models to create a 4-dimensional histogram by using a highly specialised algorithm to weight and bin the data appropriately. Before `hoki`, the only way to perform this fundamental task was to employ a code written in IDL (Interactive Data Language) and released alongside the data (see BPASS User Manual). For astronomers who do not own an IDL license, or for the large fraction of people who use Python, there was no alternative. Other complex data structures, such as Hertzsprung-Russel (HR) Diagrams, have no supporting utility code even in IDL. This situation has led to a duplication of efforts and a reproducibility concern.

With `hoki` we aim to provide the Astronomy community with an open-source Python package that streamlines the use of the BPASS data and performs complex pre-processing in the background.

## Features

The main features of the package include:

- Functionalities to automatically load BPASS outputs into Python-friendly data structures (either a `pandas.DataFrame` or a specialised object).

- Dedicated infrastructure and algorithms to handle stellar models and create custom data products that observers may need to compare to observations. This includes classes designed to contain HR diagrams and make CMDs with any combination of filters included in BPASS.

- Specialised data visualisation tools to quickly create and customize publication-ready figures.

- Standard BPASS constants, for internal `hoki` usage and advanced BPASS users.

`hoki` will remain in active development to incorporate new features as users request them. Furthermore, long term goals have been set for the implementation of more specialised functionalities, in particular SED fitting procedures to find matching BPASS models to a set of photometric, spectroscopic, or integral field unit observations. These additions will not impact the core functionalities of `hoki` described here.

In addition to the package, we have created comprehensive documentation and Jupyter notebooks to help users get started ([see documentation](#)).

## Research

`hoki` is being developed as part of a three-year project on gravitational wave counterparts and their host galaxies. Although gravitational waves are the focus of the grant funding the development of `hoki`, its use is absolutely not limited to these systems. It is designed to be versatile and usable in all applications that BPASS may be involved with. Our study of the ages and masses of the HII regions in NGC 300 relies on the use of `hoki`. In addition to streamlining the analysis, our package also allows for a transparent workflow which is easily incorporated into Jupyter notebooks that can be released alongside the peer-reviewed publication. This increases the reproducibility and accessibility of our results to the rest of the community.

## Acknowledgments

The authors are thankful for the support of the Marsden Fund Council managed through the Royal Society Te Aparangi.